\newcommand{\ket}[1]{\mbox{$ | #1 \rangle $}}
\newcommand{\bra}[1]{\mbox{$ \langle #1 | $}}

\newcommand{\beq}{\begin{eqnarray}}
\newcommand{\eeq}{\end{eqnarray}}

\documentclass[aps,pra,twocolumn,showpacs,superscriptaddress]{revtex4-1}

\usepackage[T1]{fontenc}
\usepackage[utf8]{inputenc}
\usepackage{amsmath}
\usepackage{amsfonts}
\usepackage{amsthm}
\usepackage{amssymb}
\usepackage{subeqnarray}
\usepackage{dsfont} 
\usepackage{setspace}
\usepackage{graphics}
\usepackage{url}
\usepackage{hyperref}
\usepackage{ifthen}
\usepackage{color}
\usepackage{graphicx}
\usepackage{enumitem}
\usepackage{float}

\begin{document}

\title{Non-local games and optimal steering at the boundary of the quantum set}

\author{Yi-Zheng Zhen}
\affiliation{Hefei National Laboratory for Physical Sciences at Microscale and Department of Modern Physics, University of Science and Technology of China, Hefei, Anhui 230026, People's Republic of China}
\affiliation{CAS Center for Excellence and Synergetic Innovation Center in Quantum Information and Quantum Physics, University of Science and Technology of China, Hefei, Anhui 230026, People's Republic of China}

\author{Koon Tong Goh}
\affiliation{Centre for Quantum Technologies, National University of Singapore, 3 Science Drive 2, Singapore 117543}

\author{Yu-Lin Zheng}
\affiliation{Hefei National Laboratory for Physical Sciences at Microscale and Department of Modern Physics, University of Science and Technology of China, Hefei, Anhui 230026, People's Republic of China}
\affiliation{CAS Center for Excellence and Synergetic Innovation Center in Quantum Information and Quantum Physics, University of Science and Technology of China, Hefei, Anhui 230026, People's Republic of China}

\author{Wen-Fei Cao}
\affiliation{Hefei National Laboratory for Physical Sciences at Microscale and Department of Modern Physics, University of Science and Technology of China, Hefei, Anhui 230026, People's Republic of China}
\affiliation{CAS Center for Excellence and Synergetic Innovation Center in Quantum Information and Quantum Physics, University of Science and Technology of China, Hefei, Anhui 230026, People's Republic of China}

\author{Xingyao Wu} 
\affiliation{Centre for Quantum Technologies, National University of Singapore, 3 Science Drive 2, Singapore 117543}

\author{Kai Chen} 
\affiliation{Hefei National Laboratory for Physical Sciences at Microscale and Department of Modern Physics, University of Science and Technology of China, Hefei, Anhui 230026, People's Republic of China}
\affiliation{CAS Center for Excellence and Synergetic Innovation Center in Quantum Information and Quantum Physics, University of Science and Technology of China, Hefei, Anhui 230026, People's Republic of China}

\author{Valerio Scarani}
\affiliation{Centre for Quantum Technologies, National University of Singapore, 3 Science Drive 2, Singapore 117543}
\affiliation{Department of Physics, National University of Singapore, 2 Science Drive 3, Singapore 117542}

\begin{abstract}
The boundary between classical and quantum correlations is well characterised by linear constraints called Bell inequalities. It is much harder to characterise the boundary of the quantum set itself in the space of no-signaling correlations. For the points on the quantum boundary that violate maximally some Bell inequalities, Oppenheim and Wehner [Science \textbf{330}, 1072 (2010)] pointed out a complex property: the optimal measurements of Alice steer Bob's local state to the eigenstate of an effective operator corresponding to its maximal eigenvalue. This effective operator is the linear combination of Bob's local operators induced by the coefficients of the Bell inequality, and it can be interpreted as defining a fine-grained uncertainty relation. It is natural to ask whether the same property holds for other points on the quantum boundary, using the Bell expression that defines the tangent hyperplane at each point. We prove that this is indeed the case for a large set of points, including some that were believed to provide counterexamples. The price to pay is to acknowledge that the Oppenheim-Wehner criterion does not respect equivalence under the no-signaling constraint: for each point, one has to look for specific forms of writing the Bell expressions.

%The boundary between classical and quantum correlations is well characterised by linear con-straints called Bell inequalities. It is much harder to characterise the boundary of the quantum setitself in the space of no-signaling correlations. By looking at the question from the perspective ofnon-local games based on steering, Oppenheim and Wehner [Science330, 1072 (2010)] found anintriguing property of specific points of the quantum boundary: the local state of Bob is steeredto one that saturates a local fine-grained uncertainty relation. In this paper, we extend this ob-servation to a much larger set of points, and show that recently reported counterexamples can becircumvented by exploiting the arbitrariness in the definition of a game. These results hint to thepossibility that the Oppenheim-Wehner observation holds for the whole of the quantum boundary.

\end{abstract}

\maketitle
\section{Introduction}

The violation of Bell inequalities \cite{Bell1964,Brunner2014} is one of the clearest examples in which quantum resources outperform classical ones: the outcomes of local measurements on a shared entangled quantum state cannot in general be reproduced by reading shared classical information. In other words, the \textit{quantum set} of probability distributions is strictly larger than the \textit{(classical) local set}. It is also known that shared entanglement cannot be used at a later time to exchange a message, but quantum theory does not achieve all the possible \textit{no-signaling} probability distribution \cite{Popescu1994}.

The resources ``shared quantum states'' thus hang somehow between two easily formulated types of resources. This observation has triggered the effort to find a physical or information-theoretical principle that would identify the quantum set. Some of these tentative principles are device-independent: they can be formulated at the level of probability distributions \cite{Pawlowski2009,Navascues2009}; they failed to reach the quantum set \cite{Navascues2015}. Others rely on the quantum formalism and extrapolate from it. The observation made by Oppenheim and Wehner (OW) \cite{oppenheim2010}, which is the focus of this paper, belongs to the latter category. In a nutshell, OW recast Bell inequalities in terms of steering from Alice to Bob, in a way that highlights that violation can come from two contributions: the capacity of Alice to steer Bob's state, and the ``certainty'' of the statistics of Bob's different measurements. Then, for several examples including the iconic Clauser-Horne-Shimony-Holt (CHSH) inequality \cite{CHSH}, they proved that the point of maximal quantum violation is characterized by both full steerability and an amount of certainty as high as allowed by some uncertainty relations.

A link between bipartite nonlocality and local uncertainty relations is definitely an observation worth closer scrutiny. In particular, it is natural to ask whether the property holds beyond the examples studied in Ref.~\cite{oppenheim2010}, and ultimately perhaps for the whole of the quantum boundary. A few months ago, Ramanathan and coworkers answered negatively to the latter conjecture, by producing Bell expressions for which the OW property does not hold \cite{Ramanathan2015}. Though correct, their conclusion overlooks a subtle feature of the OW property that was known by some but had never been highlighted in a publication. We need to explain this feature to motivate our contribution (more details will be given in \ref{ssnotequiv} below).

If $I\leq I_L$ is a Bell inequality, for every constant $k$ it is clear that $I+k\leq I_L+k$ is an equivalent way of writing the same inequality (in particular, the point on the quantum boundary that reaches the quantum maximum $I_Q$ is going to be the same). Now, the constant $k$ can be written in various ways exploiting the normalisation of probabilities and the no-signaling constraints. For instance, it is known that the so-called CH inequality \cite{CH} is equivalent to CHSH for no-signaling correlations; and everyone who has worked with the Collins-Linden-Gisin-Massar-Popescu (CGLMP) inequalities knows that various authors adopt different ways of expressing them \cite{CGLMP,zohren2008,Scarani2006}. The algorithmic classification of Bell inequalities by Rosset and coworkers \cite{DENIS} makes a systematic use of this equivalence. Now, it turns out that \textit{the OW way of dealing with Bell inequalities does not respect this equivalence}: if the OW property holds for a Bell expression $I$, it may well not hold for an equivalent Bell expression $I+k$ if the constant $k$ is written in a way that exploits the no-signaling constraint.

On the one hand, this features makes the OW approach significantly weaker than one might have hoped. On the other hand though, to prove that the OW property does not hold for some points on the quantum boundary, it is not sufficient to work with specific Bell expressions: one should prove that, given a point on the quantum boundary, OW holds for \textit{no} Bell expression that is maximised by that point. In short, since the OW property is at best a ``there exist'' statement, its negation must take the form of a ``for all'' statement.

In this paper, we are going to show that the OW link holds for several known examples of points on the quantum boundary, including those of the alleged counterexamples just mentioned. No counterexample having been found, it remains an open conjecture whether the OW property holds on the whole of the quantum boundary.

\section{Methodology}
\label{secmethod}

\subsection{Definitions and notations}

Consider a bipartite $(m_A,m_B,n_A,n_B)$-scenario: Alice's measurements are indexed by $x\in\{1,...,m_A\}$ and their outcomes are labeled $a\in\{1,...,n_A\}$; Bob's measurements are indexed by $y\in\{1,...,m_B\}$ and their outcomes are labeled $b\in\{1,...,n_B\}$. A probability point $P$ will be described by all the $P(ab|xy)$. A Bell inequality, denoted by $I(P)$, can be written as a linear sum of conditional probabilities upper bounded by its maximum value achievable by local realistic resources, $I_L$.
\begin{equation}
I(P):=\sum_{abxy} V_{abxy} P(a,b|x,y) \leq I_L
\label{bell}
\end{equation}
For a given Bell inequality, the values of $V_{abxy}$ are not uniquely defined. Hence, different $V_{abxy}$, or Bell expressions, are equivalent Bell inequalities which are maximally violated by the same point on the quantum boundary. By rewriting the left-hand side of equation \eqref{bell}:
\begin{align}
\sum_{abxy} V_{abxy} P(a,b|x,y) &= \sum_{ax} P(a|x) \sum_{by} V_{abxy} P(b|y,a,x) \nonumber\\
&\equiv \sum_{ax} P(a|x) \big<\hat{B}(x,a)\big>_{\rho_{B|x,a}}
\end{align}
where $\hat{B}(x,a)=\sum_{by}V_{abxy}\Pi^y_b$. Denoting $\lambda_{(x,a)}$ the largest eigenvalue of $\hat{B}(x,a)$, it obviously holds
$\big<\hat{B}(x,a)\big>_{\rho_{B|x,a}} \leq \lambda_{(x,a)}$. Inspired by the approach of Oppenheim and Wehner \cite{oppenheim2010}, we call \textit{OW-game} a Bell expression for which these inequalities are saturated at the boundary of the quantum set:
\begin{equation}\label{owcriterion}
I(P)=I_Q\;\Longrightarrow\;\big<\hat{B}(x,a)\big>_{\rho_{B|x,a}} = \lambda_{x,a} \;\; \forall x,a
\end{equation}
where $I_Q$ is the maximum value of $I(P)$ achievable by quantum resources. In words: in an OW-game, with her optimal measurements Alice steers Bob's state precisely to the eigenvector of the effective operator $B(x,a)$, for all inputs $x$ and outputs $a$. Oppenheim and Wehner proved that several XOR games, including the one based on the CHSH inequality, are OW-games.

Apart from the contents of subsections \ref{sscglmpd} and \ref{ssmermin}, the results of this paper are obtained in the $(2,2,2,2)$ Bell scenario, so we introduce a convenient notation. Labelling $x,y\in\left\{ 0,1\right\}$ and $a,b\in\left\{ 0,1\right\}$, we represent Bell expressions as tables
\begin{equation}
I=
\begin{array}{ | c c | c c | }
  \hline 		
 V_{0000} & V_{0100} & V_{0001} & V_{0101} \\
 V_{1000} & V_{1100} & V_{1001} & V_{1101} \\ \hline
 V_{0010} & V_{0110} & V_{0011} & V_{0111} \\
 V_{1010} & V_{1110} & V_{1011} & V_{1111} \\
  \hline  
\end{array}\,.
\label{bellexp}
\end{equation}

\subsection{The OW property does not respect equivalence under no-signaling}
\label{ssnotequiv}

In the $(2,2,2,2)$ scenario, the only tight Bell inequality (facet of the local polytope) is the famous CHSH inequality \cite{CHSH}. Its maximal violation defines a single point on the quantum boundary. The basic results of the Oppenheim-Wehner paper is that \eqref{owcriterion} holds for that point. What is not explicit from the paper is that the conclusion depends on the Bell expression that is used: the XOR CHSH game is an OW-game, but other Bell expressions that define the same inequality (and are in particular maximised by the same point) may not be OW-games. This observation is the basis of all this work, so let us provide an explicit example.

We start with the CHSH XOR game $I_{CHSH}=\sum_{x,y=0}^{1}P\left(a_{x}\oplus b_{y}=xy\right)$, where $P\left(a_{x}\oplus b_{y}=xy\right)=\sum_{a}P\left(a,b=xy\oplus a|xy\right)$. It has $I_{CHSH}^L=2$ and $I_{CHSH}^Q=2+\sqrt{2}$. The state and measurements that achieve $I_Q$ can be uniquely written, up to local isometries, as $\ket{\Phi^+} =\left(\left|00\right\rangle +\left|11\right\rangle \right)/\sqrt{2}$,
and $A_{0}=\sigma_{z}$, $A_{1}=\sigma_{x}$, $B_{0}=\left(\sigma_{z}+\sigma_{x}\right)/\sqrt{2}$, $B_{1}=\left(\sigma_{z}-\sigma_{x}\right)/\sqrt{2}$  \cite{Popescu1992}. Now, consider the following re-writing:
\beq
   I_{CHSH}&=& \begin{array}{ | c c | c c | }
  \hline 		
 1 & 0 & 1 & 0 \\
 0 & 1 & 0 & 1 \\ \hline
 1 & 0 & 0 & 1 \\
 0 & 1 & 1 & 0 \\
  \hline  
\end{array}\,=\,2\,\begin{array}{ | c c | c c | }
  \hline 		
 1 & 1 & 1 & 0 \\
 0 & 1 & 1 & 1 \\ \hline
 1 & 0 & 0 & 1 \\
 1 & 1 & 0 & 0 \\
  \hline  
\end{array} \;-\; \begin{array}{ | c c | c c | }
  \hline 		
 1 & 2 & 1 & 0 \\
 0 & 1 & 2 & 1 \\ \hline
 1 & 0 & 0 & 1 \\
 2 & 1 & -1 & 0 \\
  \hline  
\end{array}\nonumber\\
&=&2I_2-3\,.
\eeq
The Bell expression $I_2$ is CGLMP$_2$ in the version of Zohren and Gill \cite{zohren2008}, which was already known to be equivalent to CHSH for no-signaling $P$'s. Indeed, let's prove that the rightmost table is a complicated way of writing the constant $k=3$ for no-signaling $P$'s. The top left block is $P(00|00)+2P(01|00)+P(11|00)=P_A(0|0)+P_B(1|0)$. Treating the two off-diagonal blocks similarly, we find that the table represents $[P_A(0|0)+P_B(1|0)]+[P_A(1|0)+P_B(0|1)]+[P_A(1|1)+P_B(0|0)]+P(01|11)-P(10|11) = 2+P_A(1|1)+P_B(0|1)+P(01|11)-P(10|11)$. Again because of no-signaling, one has $P_A(1|1)=P(10|11)+P(11|11)$ and $P_B(0|1)=P(00|11)+P(10|11)$, which proves the claim.

It follows from these observations that $I_2^Q$ is obtained for the same point on the quantum boundary that gives $I_{CHSH}^Q$, which as we said is achievable only with the states and measurements written above. Then, given the operators, the bounds $\langle B(x,a)\rangle \leq\lambda_{(x,a)}$ of $I_2$ are given in terms of the $P(b|y,a_x)$ by
\begin{align}
P(0|0,0_0)+P(1|0,0_0)+P(0|1,0_0) &\leq \lambda_{2(0,0)},\nonumber \\
P(1|0,1_0)+P(0|1,1_0)+P(1|1,1_0) &\leq \lambda_{2(0,1)},\nonumber \\
%P(0|1\;0_0) &\leq \lambda_{2(0,0)},\nonumber \\ 
%P(1|0\;1_0) &\leq  \lambda_{2(0,1)}, \nonumber \\
P(0|0,0_1) + P(1|1,0_1) &\leq \lambda_{2(1,0)},\nonumber\\
P(0|0,1_1)+P(1|0,1_1) &= 1,\nonumber
\end{align}
where $\lambda_{2(0,0)}=\lambda_{2(0,1)}=2$ and $\lambda_{2(1,0)}=1+\frac{1}{\sqrt{2}}$ \footnote{Notice that conditions like $P(0|0,0_0)+P(1|0,0_0)=1$ hold by definition, so one could just have written the first line as $P(0|1,0_0) \leq \lambda'_{2(0,0)}$ where $\lambda'_{2(0,0)}=\lambda_{2(0,0)}-1$; and similarly for the second line.}. But for the state under consideration, $\langle B(0,a)\rangle=\frac{3}{2}+\frac{1}{2\sqrt{2}}<2=\lambda_{2(0,a)}$ for both $a=0,1$. Hence, the CGLMP$_2$ game $I_2$ is not an OW-game. This case study shows that different Bell expressions may behave differently on the OW characterization.

\subsection{Transformations that represent equivalence under no-signaling}

As explained with the example of CHSH and CGLMP$_2$, we are going to look for alternative Bell expressions of the same inequality obtained by adding a constant term $k$, and check if at least one of them defines an OW-game.

In our notation, a Bell expression of a Bell inequality $I$ is represented by the table \eqref{bellexp} --- we keep the discussion in the (2,2,2,2) Bell scenario, but the generalisation is obvious. First notice that tables of the type
\begin{equation}
k=
\begin{array}{ | c c | c c | }
  \hline 		
 k & k & 0 & 0 \\
 k & k & 0 & 0 \\ \hline
 0 & 0 & 0 & 0 \\
 0 & 0 & 0 & 0 \\
  \hline  
\end{array}=
\begin{array}{ | c c | c c | }
  \hline 		
 0 & 0 & 0 & 0 \\
 0 & 0 & 0 & 0 \\ \hline
 k & k & 0 & 0 \\
 k & k & 0 & 0 \\
  \hline  
\end{array} = 
\begin{array}{ | c c | c c | }
  \hline 		
 0 & 0 & k & k \\
 0 & 0 & k & k \\ \hline
 0 & 0 & 0 & 0 \\
 0 & 0 & 0 & 0 \\
  \hline  
\end{array}=
\begin{array}{ | c c | c c | }
  \hline 		
 0 & 0 & 0 & 0 \\
 0 & 0 & 0 & 0 \\ \hline
 0 & 0 & k & k \\
 0 & 0 & k & k \\
  \hline  
\end{array}\,,
\end{equation}
for a real number $k$, and convex combinations thereof, do indeed represent the constant $k$ due to the normalisation constraint $\sum_{a,b}P(a,b|x,y)=1$ $\forall x,y$. This representation of a constant is pretty trivial and indeed one can check that the OW character of a Bell expression is not changed by adding $k$ expressed in this way. For instance, starting from a Bell expression $I$, one can always construct $I'$ with the same OW character such that all the $V_{abxy}$ are non-negative.

If we now enforce the no-signalling constraint for Bob $P(b|x=0,y)=P(b|x=1,y)=P(b|y)$, we obtain less trivial representations of the same constant:
\begin{equation}
k=
\begin{array}{ | c c | c c | }
  \hline 		
 k & 0 & 0 & 0 \\
 k & 0 & 0 & 0 \\ \hline
 0 & k & 0 & 0 \\
 0 & k & 0 & 0 \\
  \hline  
\end{array}=
\begin{array}{ | c c | c c | }
  \hline 		
 0 & k & 0 & 0 \\
 0 & k & 0 & 0 \\ \hline
 k & 0 & 0 & 0 \\
 k & 0 & 0 & 0 \\
  \hline  
\end{array} = 
\begin{array}{ | c c | c c | }
  \hline 		
 0 & 0 & k & 0 \\
 0 & 0 & k & 0 \\ \hline
 0 & 0 & 0 & k \\
 0 & 0 & 0 & k \\
  \hline  
\end{array}=
\begin{array}{ | c c | c c | }
  \hline 		
 0 & 0 & 0 & k \\
 0 & 0 & 0 & k \\ \hline
 0 & 0 & k & 0 \\
 0 & 0 & k & 0 \\
  \hline  
\end{array}\,,
\end{equation}
and convex combinations thereof. Similarly, enforcing the no-signaling constraint for Alice $P(a|x,y=0)=P(a|x,y=1)=P(a|x)$, we have the additional representations
\begin{equation}
k=
\begin{array}{ | c c | c c | }
  \hline 		
 k & k & 0 & 0 \\
 0 & 0 & k & k \\ \hline
 0 & 0 & 0 & 0 \\
 0 & 0 & 0 & 0 \\
  \hline  
\end{array}=
\begin{array}{ | c c | c c | }
  \hline 		
 0 & 0 & k & k \\
 k & k & 0 & 0 \\ \hline
 0 & 0 & 0 & 0 \\
 0 & 0 & 0 & 0 \\
  \hline  
\end{array}=
\begin{array}{ | c c | c c | }
  \hline 		
 0 & 0 & 0 & 0 \\
 0 & 0 & 0 & 0 \\ \hline
 k & k & 0 & 0 \\
 0 & 0 & k & k \\
  \hline  
\end{array}
=
\begin{array}{ | c c | c c | }
  \hline 		
 0 & 0 & 0 & 0 \\
 0 & 0 & 0 & 0 \\ \hline
 0 & 0 & k & k \\
 k & k & 0 & 0 \\
  \hline  
\end{array}\,,
\end{equation}
and convex combinations thereof. Such rewritings based on no-signaling may change the OW character of a Bell expression, as it happened in the CHSH example above.

Having come to terms with this flexibility, it is convenient to recast the same information in \textit{difference tables} 
\begin{equation}\label{diff}
D\,=\,\begin{array}{|c|c|}
  \hline D_{00}&D_{01}\\
  \hline
  D_{10}&D_{11}\\\hline
 \end{array}
 \end{equation}
with
\begin{equation}
 D_{xy}=\begin{array}{|ccc|}\hline &V_{00xy}-V_{01xy}& \\V_{00xy}-V_{10xy} &&V_{01xy}-V_{11xy}\\&V_{10xy}-V_{11xy}&\\\hline\end{array}\,.
\end{equation}
This representation is handy because the transformations allowed by no-signaling take a very simple form. Indeed, two difference tables are equivalent under no-signaling if and only if there exist $\alpha, \beta, \gamma, \delta \in \mathds{R}$ such that $D'=D+\Delta(\alpha, \beta, \gamma, \delta)$ with
\begin{align}
\Delta(\alpha, \beta, \gamma, \delta)\,=\, \begin{array}{ | c c c | c c c | }  \hline 
&+\alpha&&&+\beta& \\ +\gamma &&+\gamma&-\gamma& &-\gamma \\ &+\alpha&&&+\beta& \\ \hline
&-\alpha&&&-\beta& \\ +\delta &&+\delta&-\delta& &-\delta \\ &-\alpha&&&-\beta& \\
\hline  \end{array}\,.
\end{align}

In particular, if a difference table $D_k$ represents a constant $k$ under the no-signaling constraints for Alice and Bob, there must exist $\alpha, \beta, \gamma, \delta$ such that
\begin{equation}\label{condconst}
D'_k = D_k+\Delta(\alpha, \beta, \gamma, \delta) = 
\begin{array}{ | c c c | c c c | }
  \hline 		
  &0&&&0& \\
 0 &&0&0&&0 \\
 &0&&&0& \\ \hline
&0&&&0& \\
0&&0&0&&0 \\
&0&&&0& \\ 
  \hline  
\end{array}\,.
\end{equation}
%Hence, an expression is a constant if and only if there exist $\alpha, \beta, \gamma, \delta \in \mathds{R}$ such that:
%\begin{align}&V_{0000}-V_{0100}+\alpha=V_{1000}-V_{1100}+\alpha \label{condconst}\\=&V_{0010}-V_{0110}-\alpha=V_{1010}-V_{1110}-\alpha\nonumber\\=&V_{0001}-V_{0101}+\beta=V_{1001}-V_{1101}+\beta\nonumber\\=&V_{0011}-V_{0111}-\beta=V_{1011}-V_{1111}-\beta\nonumber\\=&V_{0000}-V_{1000}+\gamma=V_{0100}-V_{1100}+\gamma\nonumber\\=&V_{0001}-V_{1001}-\gamma=V_{0101}-V_{1101}-\gamma\nonumber\\=&V_{0010}-V_{1010}+\delta=V_{0110}-V_{1110}+\delta\nonumber\\=&V_{0011}-V_{1011}-\delta=V_{0111}-V_{1111}-\delta=0\nonumber\end{align}

\subsection{Checking for OW-games}
\label{ssnec}

Having introduced the context, we can finally explain how one can look for an OW-game.

First notice that even the verification of the OW criterion for a given Bell expression is not trivial \textit{a priori}. Indeed, while the property of ``being at the quantum boundary'' is determined by the Bell expression alone, checking the OW criterion involves finding the states and operators that realize the quantum point $P$. For a generic Bell expression, it is not known how to find such a quantum realisation; and even once one is found, there is no guarantee that it is unique. If $P$ could be obtained with inequivalent realisations of the state and the measurements, one would have to say whether saturation of \eqref{owcriterion} holds for all realisations, or it is enough that it holds for one. As it turns out, for all the cases explicitly studied so far, a $P$ on the quantum boundary can be obtained by a unique choice of the state and the measurements, up to local isometries (``\textit{self-testing}''). The independent conjecture that self-testing holds on the whole quantum boundary is interesting in its own right, but we don't address it here.

Having clarified how the OW criterion is going to tested, we need to move one step back and explain how one can try and guess a Bell expression that is a candidate for OW-game, given all the freedom allowed by the equivalence under no-signaling. The heuristic method we found consists in enforcing first some necessary conditions. Indeed, it is clear that the OW criteria \eqref{owcriterion} can only be satisfied if $\hat{B}(x,a)$ is diagonal in the basis which $\rho_{B|x,a}$ is diagonal. This condition imposes several constraints on $V_{abxy}$, that largely restrict the candidate Bell expressions. The remaining ones can then be tested directly. In Appendix \ref{appc} we describe in greater detail how these constraints are used in the case of self-testing points in the $(2,2,2,2)$ Bell scenario.

\section{Points with OW-game}
\label{secow}

We present now the points on the quantum boundary for which we have found an OW-game. For the $(2,2,2,2)$ Bell scenario, we first discuss the two points that allegedly provided counterexamples (\ref{ssravi}), then two whole families of points (\ref{ssfamilies}). Then we present one point in the $(2,2,d,d)$ Bell scenario, the one that violates maximally the CGLMP$_d$ inequality (\ref{sscglmpd}). Finally, one point in a three-partite scenario, the one that violates maximally the Mermin inequality, together with the suitable definition of steering in the multipartite case (\ref{ssmermin}).

\subsection{Alleged counterexamples}
\label{ssravi}

Ramanathan and coworkers \cite{Ramanathan2015} provided two games that are not OW-games. As we know by now, this is not sufficient to prove that there is no OW-game for the corresponding points on the quantum boundary --- and as it turns out, there is.

For the first point, we use the family of Bell expressions
\begin{equation}
I_{c1}(\Gamma)=
\begin{tabular}{ | c c | c c | }
  \hline 		
 1 & 0 & 0 & 0 \\
 0 & 0 & 1 & 0 \\ \hline
 0 & 1 & 0 & 1 \\
 1 & 0 & 0 & 0 \\
  \hline  
\end{tabular} +
\begin{tabular}{ | c c | c c | }
  \hline 		
 $\Gamma$ & 1 & $\Gamma$ & 1 \\
 $\Gamma$ & 1 & $\Gamma$ & 1 \\ \hline
 0 & $\Gamma-1$ & 0 & $\Gamma-1$ \\
 0 & $\Gamma-1$ & 0 & $\Gamma-1$ \\
  \hline  
  \end{tabular}=I_{c1}(0)+2\Gamma\,.
\end{equation} The fact that the rightmost table is equal to $k=2\Gamma$ for no-signaling $P$'s can be checked with the tools described above. The bounds $\langle B(x,a)\rangle \leq\lambda_{(x,a)}$ now read
\begin{align}
\Gamma P(0|0,0_0) + (1-\Gamma) P(1|1,0_0) &\leq \lambda_{c1(0,0)},\nonumber \\ 
(1-\Gamma)P(1|0,1_0) + \Gamma P(0|1,1_0) &\leq  \lambda_{c1(0,1)}, \nonumber \\
P(1|0,0_1) + P(1|1,0_1) &\leq \lambda_{c1(1,0)},\nonumber \\ 
(2-\Gamma)P(0|0,1_1) + (1-\Gamma) P(0|1,1_1) &\leq  \lambda_{c1(1,1)}, \nonumber 
\end{align}
In Ref.~\cite{Ramanathan2015} it is proved that $I_{c1}(0)$ self-tests a given two-qubit state and suitable measurements; and that the OW criteria \eqref{owcriterion} do not hold. However, using those same state and measurements, for $\Gamma \simeq 0.4648162$ we find numerically that the criteria hold, with $\lambda_{c1(0,0)}=\lambda_{c1(0,1)}\simeq0.821605$, $\lambda_{c1(1,0)}\simeq1.76759$, $\lambda_{c1(1,1)}\simeq1.89197$.

For the second point, we use the Bell expressions
\begin{equation}
I_{c2}(\Gamma)=
\begin{tabular}{ | c c | c c | }
  \hline 		
 1 & 0 & 0 & 1 \\
 0 & 1 & 1 & 0 \\ \hline
 0 & 1 & 0 & 1 \\
 1 & 0 & 0 & 0 \\
  \hline  
\end{tabular}
+
\begin{array}{ | c c | c c | }
  \hline 		
 0 & \Gamma & -\Gamma & 0 \\
 0 & \Gamma & -\Gamma & 0 \\ \hline
 \Gamma & 0 & 0 & -\Gamma \\
 \Gamma & 0 & 0 & -\Gamma \\
  \hline  
\end{array}
=I_{c2}(0)+0
\end{equation} because the table we added is equal to $k=0$ for no-signaling $P$'s. The corresponding bounds
\begin{align}
(1-\Gamma) P(0|0,0_0) + (1+\Gamma) P(1|1,0_0) &\leq \lambda_{c2(0,0)},\nonumber \\ 
(1+\Gamma)P(1|0,1_0) + (1-\Gamma) P(0|1,1_0) &\leq  \lambda_{c2(0,1)}, \nonumber \\
P(1|0,0_1) + P(1|1,0_1) &\leq \lambda_{c2(1,0)},\nonumber \\ 
(1+\Gamma)P(0|0,1_1) +\Gamma P(0|1,1_1) &\leq  \lambda_{c2(1,1)}, \nonumber 
\end{align} are not saturated for $\Gamma=0$, as proved in Ref.~\cite{Ramanathan2015}; but they are for $\Gamma \simeq 0.5601320$, in which case $\lambda_{c2(0,0)}=\lambda_{c2(0,1)}=\lambda_{c2(1,1)}\simeq1.84450$, $\lambda_{c2(1,0)}\simeq1.64649$.

\subsection{Families of points}
\label{ssfamilies}

Even for the $(2,2,2,2)$-scenario, we do not know a complete parametrisation of the quantum boundary. The most famous family of points is the \textit{three-parameter family} that describes the slice with unbiased marginals $P(a|x)=P(b|y)=\frac{1}{2}$. The boundary is known to be given by $\sum_{x,y}(-1)^{xy}\textrm{Arcsin}(E_{xy})=\pi$ with $E_{xy}=P(a=b|xy)-P(a\neq b|xy)$, or suitable permutations of the settings and the outcomes \cite{Cirelson1980,Landau1988}. The points on these boundaries are also those that self-test $\ket{\Phi^+}$ in the $(2,2,2,2)$-scenario \cite{Wang2015}. Now, for $E_{xy}\equiv\cos\alpha_{xy}\neq \pm 1$, the inequality that describes the tangent to each of these points can be cast as the game \cite{Miller2012,Wang2015}
\begin{equation}
I_{\vec{E}}=
\begin{array}{ | c c | c c | }
  \hline 		
 \frac{1}{\sin\alpha_{00}} & 0 & \frac{1}{\sin\alpha_{01}} & 0 \\
 0 & \frac{1}{\sin\alpha_{00}} & 0 & \frac{1}{\sin\alpha_{01}} \\ \hline
 \frac{1}{\sin\alpha_{10}} & 0 & -\frac{1}{\sin\alpha_{11}} & 0 \\
 0 & \frac{1}{\sin\alpha_{10}} & 0 & -\frac{1}{\sin\alpha_{11}} \\
  \hline  
\end{array}\,.
\end{equation}
This is a weighted XOR game, the non-zero $V_{abxy}$ being different for different $(x,y)$. We checked numerically that the OW criteria \eqref{owcriterion} hold by sampling 156849 such points at random. 

The other family that we consider is the \textit{one-parameter} family of the points that violate maximally one of the \textit{tilted CHSH inequalities} $\alpha E^A_0+E_{00}+E_{01}+E_{10}-E_{11}\leq 2+\alpha$ where $E^A_0=P_A(0|0)-P_A(1|0)$ and $\alpha \in [0,2]$ \cite{acin2002}. Each of these points self-tests a corresponding partially entangled qubit state  $\ket{\psi(\theta)}=\cos \theta \ket{00} + \sin \theta \ket{11}$  with $\alpha = 2/\sqrt{1+2\tan^2 2\theta}$, for the measurements $A_0 = \sigma_z$,
$A_1 = \sigma_x$, $B_0 = \cos \mu \sigma_z + \sin \mu \sigma_x$ and $B_1 = \cos \mu \sigma_z - \sin \mu \sigma_x$ where $\tan \mu = \sin 2\theta$ \cite{Yang2013, Bamps2015}. For these points, we work with the family of Bell expressions
\begin{equation}
I_{\alpha}(\Gamma)=
\begin{array}{ | c c | c c | }
  \hline 		
 1+\alpha & \alpha & 1 & 0 \\
 0 & 1 & 0 & 1 \\ \hline
 1 & 0 & 0 & 1 \\
 0 & 1 & 1 & 0 \\
  \hline  
\end{array}
-
\Gamma\begin{array}{ | c c | c c | }
  \hline 		
 0 & -\cos2\theta & 0 & -\cos2\theta \\
 0 & -\cos2\theta & 0 & -\cos2\theta \\ \hline
 \sin^2\theta &  \cos^2\theta &   \sin^2\theta & \cos^2\theta \\
 \sin^2\theta &  \cos^2\theta &  \sin^2\theta & \cos^2\theta \\
  \hline  
\end{array}
\end{equation}
The rightmost table is $k=2\sin^2\theta$ for no-signaling $P$'s, so the local bound is $I^L_\alpha(\Gamma)=2+\alpha-2\Gamma\sin^2\theta$. The case $\Gamma=0$ is the game that one would naturally write down from the inequality as stated, but it can be checked that it is not an OW-game for any $\alpha\in\left(0,2\right]$ \footnote{For the state and the measurements that define the point at the quantum boundary, one finds $\big< B(1,b)\big>=1+(1-\cos 4\theta)/(\sqrt{6-2\cos 4\theta})$ for $b=0,1$. This is strictly smaller than $\lambda_{(1,b)}=1+|\sin 2\theta|\sqrt{\frac{2}{3-\cos 4\theta}}$ for $\alpha\in\left(0,2\right]$.}.
However, $\Gamma=1$ provides an OW-game. In this case, the bounds $\langle B(x,a)\rangle\leq\lambda_{(x,a)}$ are given by
\begin{align}
P(0|0,0_0) + P(0|1,0_0) &\leq \lambda_{\alpha(0,0)},\nonumber \\ 
P(1|0,1_0) + P(1|1,1_0) &\leq  \lambda_{\alpha(0,1)}, \nonumber \\
\cos^2\theta P(0|0,0_1) + \sin^2\theta P(1|1,0_1) &\leq \lambda_{\alpha(1,0)},\nonumber \\ 
\sin^2\theta P(1|0,1_1) + \cos^2\theta P(0|1,1_1) &\leq  \lambda_{\alpha(1,1)}, \nonumber 
\end{align}
with $\lambda_{\alpha(0,0)}=1+\sqrt{\frac{2}{3-\cos 4\theta}}$, $\lambda_{\alpha(0,1)}=\frac{1-\cos 4\theta }{3-\cos 4\theta-\sqrt{6-2\cos 4\theta}}$ and $ \lambda_{\alpha(1,0)}= \lambda_{\alpha(1,1)}=\frac{1}{2}+\frac{1}{\sqrt{6-2\cos4\theta}}$.

Interestingly, even if these Bell expressions are asymmetric between Alice and Bob, they can be used to steer in the either direction. We have just presented the steering from Alice to Bob. That from Bob to Alice, the OW criteria is given by $\left<B'(y,b)\right>=\lambda'_{(y,b)}\; \forall y,b$ where $B'(y,b)=\sum_{yb}V_{abxy}\Pi^x_a$ and $\lambda'_{(y,b)}$ is the largest eigenvalue of $B'(y,b)$.

From the tilted CHSH inequalities, we can write down another family of Bell expressions, denoted by $I'_\alpha$, which is given by:

\begin{equation}
I'_{\alpha}(\Gamma)=
\begin{array}{ | c c | c c | }
  \hline 		
 1+\alpha & \alpha & 1 & 0 \\
 0 & 1 & 0 & 1 \\ \hline
 1 & 0 & 0 & 1 \\
 0 & 1 & 1 & 0 \\
  \hline  
\end{array}
+
\Gamma\begin{array}{ | c c | c c | }
  \hline 		
 -\alpha & -\alpha & 0 & 0 \\
 X_1-\alpha & X_1-\alpha & -X_1 & -X_1 \\ \hline
 0 &  0 &  1 & 1 \\
 X_2 &  X_2 &  1-X_2 & 1-X_2 \\
  \hline  
\end{array}
\end{equation}
where $X_1=\frac{2-\Lambda_+-\Lambda_-}{\Lambda_+-\Lambda_-}$, $X_2=\frac{2\Lambda_+\Lambda_--\Lambda_+-\Lambda_-}{\Lambda_+-\Lambda_-}$ and $\Lambda_\pm=\frac{2\sin^22\theta}{1-2\sin^2\theta\pm\sqrt{1+\sin^22\theta}}$. Similarly, $I'_\alpha(\Gamma)$ is an OW-game for the case $\Gamma=1$ but not $\Gamma=0$. In the case $\Gamma=1$, the bounds $\left<B(y,b)\right>\leq\lambda_{(y,b)}$ are given by
\begin{align*}
P(0|0,0_0)+(X_1-\alpha)P(1|0,0_0)&\\
+P(0|1,0_0)+X_2P(1|1,0_0)&\leq\lambda'_{\alpha(0,0)},\\ (X_1+1-\alpha)P(1|0,1_0)+(X_2+1)P(1|1,1_0)&\leq\lambda'_{\alpha(0,1)},\\
P(0|0,0_1)-X_1P(1|0,0_1)+P(0|1,0_1)&\\
+(2-X_2)P(1|1,0_1)&\leq\lambda'_{\alpha(1,0)},\\
(1-X_1)P(1|0,1_1)+2P(0|1,1_1)&\\
+(1-X_2)P(1|1,1_1)&\leq\lambda'_{\alpha(1,1)}.
\end{align*}
For the steering scenario of Bob to Alice, the probabilities written above are $P(a|x\;b_y)$.

A final remark: we have also explored a third family of points, those that violate maximally the ``Hardy inequalities'' introduced by by Man{\v{c}}inska and Wehner \cite{Mancinska2014}. We have strong numerical evidence of both the fact that a sample of these points are self-testing and that one can construct OW-games for each of them. We don't think that this paper will be significantly improved by a detailed presentation of these optimisations as they stand.

\subsection{OW-games with more outcomes: maximal violation of CGLMP$_d$}
\label{sscglmpd}

In this section, we leave the $(2,2,2,2)$-scenario to discuss one point with OW-game in the $(2,2,d,d)$-scenario, for any $d\geq 3$. Concretely, we consider the points that violate maximally each of the CGLMP$_d$ inequalities. The conclusions of this subsection rely on the conjecture that the maximal quantum violation is indeed achieved by the points constructed below (proved up to numerical precision for $d\leq 8$, see Table 1 of \cite{Navascues2008}). We have also to warn that the self-testing of the states and measurements has been proved so far only for $d=3$ \cite{Yang2013}.

We denote $x,y\in\left\{ 0,1\right\} $ as the measurement settings and $a,b\in\left\{ 0,1,\dots,d-1\right\} $ as the outcomes
for Alice and Bob respectively. In this scenario, the Collins-Gisin-Linden-Massar-Popescu (CGLMP) inequalities \cite{CGLMP} are a class of facets. Maybe the most compact way of writing the CGLMP inequality is that of Zohren and Gill: \beq
I_{d}&=&P\left(a_{0}\leq b_{0}\right)+P\left(a_{0}\geq b_{1}\right)\nonumber\\&&+P\left(a_{1}\geq b_{0}\right)+P\left(a_{1}<b_{1}\right)\leq 3,\label{eq:CGLMPdineq}
\eeq
where $P\left(a_{x}\leq b_{y}\right)=\sum_{a\leq b}P\left(ab|xy\right)$
\cite{zohren2008}. As it happened for $d=2$, this form is not an OW-game (see Appendix \ref{appzg3} for the explicit proof in the case $d=3$). An OW-game based on CGLMP$_3$ was constructed in Ref.~\cite{Ramanathan2015}:
$G_{3}=\sum_{x,y=0}^{1}2P\left(a_{x}=b_{y}-xy\right)+P\left(a_{x}=b_{y}+x+y-2\right)$, where
$P\left(a_{x}=b_{y}+\Delta\right)=\sum_{k=0}^{d}P\left(a=k,b=\left(k-\Delta\right)\mathrm{mod}\;d|xy\right)$. By inspection, one finds that $G_3=3I_3-3$. 

This construction can be generalised to high dimensional case. Now consider the non-local game
\begin{scriptsize}
\begin{eqnarray}
G_{d} & = & \sum_{xy}\sum_{\Delta=0}^{d-1}\Delta P\left(a-b=\left(-1\right)^{x+y}\left(\Delta+1\right)-xy|xy\right)\nonumber \\
 & = & \begin{array}{|cccc|cccc|}
\hline  d-1 & d-2 & \cdots & 0 & d-1 & 0 & \cdots & d-2\\
 0 & d-1 & \cdots & 1 & d-2 & d-1 & \ddots & \vdots\\
 \vdots & \ddots & \ddots & \vdots & \vdots & \ddots & \ddots & 0\\
 d-2 & \cdots & 0 & d-1 & 0 & \cdots & d-2 & d-1\\
\hline  d-1 & 0 & \cdots & d-2 & 0 & d-1 & \cdots & 1\\
 d-2 & d-1 & \ddots & \vdots & 1 & 0 & \ddots & \vdots\\
 \vdots & \ddots & \ddots & 0 & \vdots & \ddots & \ddots & d-1\\
 0 & \cdots & d-2 & d-1 & d-1 & \cdots & 1 & 0 \\\hline
\end{array}.
\label{eq:CGLMPd-OW}
\end{eqnarray}
\end{scriptsize}
$G_d$ is also a weighted XOR game. It's easy to check that $G_2$ is the CHSH-XOR game \cite{oppenheim2010} and $G_{3}$ is the game presented in Ref.~\cite{Ramanathan2015}. In fact, $G_d=dI_d-3$ holds for all $d$, so $G_d$ is a CGLMP$_d$ game. It can be shown that this is an OW-game for all $d$, under the conjecture mentioned above on the form of the optimal measurements. We leave the proof of this statement to Appendix \ref{cglmpd}.

\subsection{Multipartite example: Maximal Violation of Mermin Inequality}
\label{ssmermin}

All discussions made on OW criteria in the literature focus on bipartite Bell scenarios, which is not surprising because the idea of steering is somehow naturally bipartite. However, a multipartite generalisation of steering has been introduced. Following the work of Cavalcanti and coworkers \cite{Cavalcanti2015}, in the tripartite Bell scenario, one can distinguish two types of steering: (i) 1 black box that steers to 2 characterised devices and (ii) 2 black boxes that steer to 1 characterised device. Each type of steering would give rise to different sets of OW criteria, namely:
\begin{align}
&\text{(i):}\;\; \big<\hat{B}(x,a)\big>_{\rho_{BC|x,a}} = \lambda_{(x,a)} \forall x,a\\
&\text{(ii):}\;\; \big<\hat{B}(x,y,a,b)\big>_{\rho_{C|x,y,a,b}}= \lambda_{(x,y,a,b)} \forall x,y,a,b\\
&\text{where}\;\; \hat{B}(x,a):= \sum_{bcyz}V_{abcxyz}\Pi^{y}_{b}\otimes\Pi^{z}_{c},\\
&\;\;\;\;\;\;\;\;\;\; \hat{B}(x,y,a,b):= \sum_{cz}V_{abcxyz}\Pi^{z}_{c}
\end{align}
and $\lambda_{(x,a)}$ and $\lambda_{(x,y,a,b)}$ are the largest eigenvalues of $\hat{B}(x,a)$ and $\hat{B}(x,y,a,b)$ respectively.

We study the point that violates maximally the Mermin inequality\cite{Mermin1990}
\begin{equation}
\left<A_0B_0C_0\right>-\left<A_0B_1C_1\right>-\left<A_1B_0C_1\right>-\left<A_1B_1C_0\right>\leq2
\end{equation}
When rewritten to the form of a Bell expression, we get:
\begin{equation}
I_M(P):=\sum_{abcxyz}V_{abcxyz}P(a,b,c|x,y,z)\leq3
\end{equation}
where
\begin{equation}
V_{abcxyz}=\delta_{a\oplus b\oplus c,x\vee y\vee z}\delta_{xyz,000}\delta_{xyz,011}\delta_{xyz,101}\delta_{xyz,110}
\end{equation}
The maximal quantum bound of $I_M(P)$ is given by 4 and it self-tests \cite{Pal2014} the measured quantum to be the GHZ state
\begin{equation}
\ket{\text{GHZ}}=\frac{\ket{000}+\ket{111}}{\sqrt{2}}
\end{equation}
and the measurements to be
\begin{align}
\hat{A}_0=\hat{B}_0=\hat{C}_0&=\sigma_z,\\
\hat{A}_1=\hat{B}_1=\hat{C}_1&=\sigma_y.
\end{align}

Hence, one can easily check that:
\begin{align}
&\big<\hat{B}_M(x,a)\big>_{\rho_{BC|x,a}}=\lambda_{M(x,a)}=2 \;\; \forall x,a\\
&\big<\hat{B}_M(x,y,a,b)\big>_{\rho_{C|x,y,a,b}}=\lambda_{M(x,y,a,b)}=1 \;\; \forall x,y,a,b
\end{align}
Thus, this concludes that Mermin inequality is an OW-game for both types of steering.

\section{Conclusion}
The quantum set of correlations is defined by all the $P(ab|xy)$ that can be obtained by measuring quantum states, without any constraint on the Hilbert space dimension of the underlying system. The characterization of its boundary in terms of physical or mathematical properties is still elusive. In this paper, we have shown that the Oppenheim-Wehner criteria are fulfilled by many points on the quantum boundary, including some that can't maximise any XOR game and two that were believed to provide counterexamples.

No counterexample has been found so far, which may inspire the conjecture that every point on the quantum boundary, in any scenario, has an associated OW-game. In order to test the truth of this conjecture, one would have to solve long standing problems that are of interest in themselves (and even arguably of greater interest). Indeed, in order to state the OW criteria, one needs the knowledge of the state and the measurements that realise the probability point on the quantum boundary. It is not even sure that such a point is unique: this is an open conjecture on self-testing. Even taking uniqueness for granted, it would be a breakthrough by itself, if one were able to provide the quantum realisation of the maximal violation of a generic Bell expression (even for Bell inequalities this is usually unknown).

In the context of nonlocal games, it would be interesting to study in which context the choice of a representation that is a OW-game may be an advantage. It must clearly be a situation in which the equivalence under no-signaling is not important.

\section*{Acknowledgments} Y.Z.Z., Y.L.Z. and K.C. acknowledge hospitality from CQT, Singapore, when this collaboration was started. An early discussion with Antonio Ac\'{\i}n was in the back of V.S.'s mind and triggered this project. We also thank Francesco Buscemi and Ravishankar Ramanathan for valuable discussions, and Manik Banik for highlighting typos in a previous version. 

This research is supported by the Singapore Ministry of Education Academic Research Fund Tier 3 (Grant No. MOE2012-T3-1-009); by the National Research Fund and the Ministry of Education, Singapore, under the Research Centres of Excellence programme; by the Chinese Academy of Science; and by the National Natural Science Foundation of China (Grants No.~11175170, and No.~11575174). V.S. acknowledges further support from a NUS Provost Chair grant, Y.Z.Z. and Y.L.Z. from USTC Student Scholarship and China Scholarship Council.

\bibliographystyle{aipauth4-1}
\bibliography{ref}

\begin{appendix}

\section{Enforcing necessary conditions for the $(2,2,2,2)$ Bell scenario under self-testing}
\label{appc}

In this appendix we show more explicitly how to implement the constraints discussed in section \ref{ssnec} in the case of self-testing probability distributions in the $(2,2,2,2)$ Bell scenario. In this scenario, we are guaranteed that the maximal Bell violation by a quantum resource can be achieved by a pure bipartite qubits state and projective measurements \cite{Masanes2005}. If the point is self-testing, then it does self-test a pure two-qubit state and those measurements. In particular, the steered state on Bob will be a pure qubit state.

Define now the unitary transformation $U_{x,a}$ such that
\begin{equation}
U_{x,a}\rho_{B|x,a}U^\dagger_{x,a} = \ket{0}\bra{0}\,.
\end{equation}
The projectors written in the basis where the steered state is diagonal are given in the following form:
\begin{align}
U_{x,a}\Pi^{y=0}_{b=0}U^\dagger_{x,a} &= 
\begin{pmatrix}
p_0(x,a) & q_0(x,a)\\
q_0(x,a) & 1-p_0(x,a)
\end{pmatrix}\\
U_{x,a}\Pi^{y=0}_{b=1}U^\dagger_{x,a} &= 
\begin{pmatrix}
1-p_0(x,a) & -q_0(x,a)\\
-q_0(x,a) & p_0(x,a)
\end{pmatrix}\\
U_{x,a}\Pi^{y=1}_{b=0}U^\dagger_{x,a} &= 
\begin{pmatrix}
p_1(x,a) & q_1(x,a)\\
q_1(x,a) & 1-p_1(x,a)
\end{pmatrix}\\
U_{x,a}\Pi^{y=1}_{b=1}U^\dagger_{x,a} &= 
\begin{pmatrix}
1-p_1(x,a) & -q_1(x,a)\\
-q_1(x,a) & p_1(x,a)
\end{pmatrix}
\end{align}
where $p_0(x,a)$, $q_0(x,a)$, $p_1(x,a)$ and $q_1(x,a)$ are some real numbers between 0 and 1. The necessary conditions which for a Bell expresssion to be an OW-game on $V_{abxy}$ are then
\begin{equation}
q_0(x,a)(V_{a0x0}-V_{a1x0})+q1(x,a)(V_{a0x1}-V_{a1x1}) = 0 \;\; \forall x,a\,.
\label{OWcrit}
\end{equation}
In particular, for cases where the $r(x,a):=\frac{q_0(x,a)}{q_1(x,a)}$ are well-defined for all $x,a$ pairs, an OW-game has the form of:
\begin{equation}
OW = \begin{array}{ | c c | c c | }
  \hline 		
 A & B & C & r(0,0)(A-B)+C \\
 D & E & F & r(0,1)(D-E)+F \\ \hline
 G & H & I & r(1,0)(G-H)+I \\
 J & K & L & r(1,1)(J-K)+L \\
  \hline  
\end{array} \label{OWgame}
\end{equation}
where the capital Roman alphabet letters are free variables.

In order to check whether a point may have an OW-game, we can now take any Bell expression $I$ that is maximally violated by that point, and check if there exist an OW-game such that $I-OW=k$. As discussed above, this is going to be simplest by passing in the difference representation and using equation \eqref{condconst}.

\section{The Zohren-Gill version of CGLMP$_{3}$ is not an OW-game}
\label{appzg3}

In this section, we show that $I_3$ as defined in equation (\ref{eq:CGLMPdineq}) is not an OW-game.

The bounds $\langle B(x,a)\rangle \leq\lambda_{(x,a)}$ are given by:
\begin{align}
P\left(0|0,0_{0}\right)+P\left(1|0,0_{0}\right)+P\left(2|0,0_{0}\right)+P\left(0|1,0_{0}\right)&\leq\lambda_{3\left(0,0\right)} \nonumber\\
P\left(1|0,1_{0}\right)+P\left(2|0,1_{0}\right)+P\left(0|1,1_{0}\right)+P\left(1|1,1_{0}\right)&\leq\lambda_{3\left(0,1\right)} \nonumber\\
P\left(2|0,2_{0}\right)+P\left(0|1,2_{0}\right)+P\left(1|1,2_{0}\right)+P\left(2|1,2_{0}\right)&\leq\lambda_{3\left(0,2\right)} \nonumber\\
P\left(0|0,0_{1}\right)+P\left(1|1,0_{1}\right)+P\left(2|1,0_{1}\right)&\leq\lambda_{3\left(1,0\right)} \nonumber\\
P\left(0|0,1_{1}\right)+P\left(1|0,1_{1}\right)+P\left(2|1,1_{1}\right)&\leq\lambda_{3\left(1,1\right)} \nonumber\\
P\left(0|0,2_{1}\right)+P\left(1|0,2_{1}\right)+P\left(2|0,2_{1}\right)&\leq\lambda_{3\left(1,2\right)} \nonumber
\end{align}
Since the maximal violation
of the CGLMP$_{3}$ inequality is self-testing \cite{Yang2013}, the optimal state and
measurements to violate the CGLMP$_{3}$ inequality are unique up to local isometries. The optimal state is given by $\left|\psi_{\gamma}\right\rangle =\left(\left|00\right\rangle +\gamma\left|11\right\rangle +\left|22\right\rangle \right)/\sqrt{2+\gamma^{2}}$
where $\gamma=\frac{\sqrt{11}-\sqrt{3}}{2}$ \cite{acin2002}, while
the optimal measurements are described in equation (\ref{eq:CGLMPd-measu}). Hence, it is possible to study the the inequalities $\langle B(x,a)\rangle\leq\lambda_{(x,a)}$ at the point of maximal CGLMP$_{3}$ violation. Table~\ref{Table-CGLMP3} shows the values of $\langle B(x,a)\rangle$ and $\lambda_{3(x,a)} \; \forall x,a$ when $I_3(P)=I_Q$: since they are different, the non-local game $I_3$ does not exhibit the OW property. 

\begin{table}[h]
\caption{$B(x,a)$ and $\lambda_{3\left(x,a\right)}$ of
the CGLMP$_{3}$ game $I_3$}\label{Table-CGLMP3}
\begin{tabular}{|c|c|c|c|c|c|c|}
\hline 
$\left(x,a\right)$ & $\left(0,0\right)$ & $\left(0,1\right)$ & $\left(0,2\right)$ & $\left(1,0\right)$ & $\left(1,1\right)$ & $\left(1,2\right)$\tabularnewline
\hline 
$\lambda_{3\left(x,a\right)}$ & 2 & 2 & 2 & 1.7454 & 1.7454 & 1\tabularnewline
\hline 
$\langle B(x,a)\rangle$ & 1.8083 & 1.8407 & 1.8083 & 1.7287 & 1.7287 & 1\tabularnewline
\hline 
\end{tabular}
\end{table}

\section{OW-games for the maximal violation of CGLMP$_{d}$}
\label{cglmpd}

In this Appendix, we provide the explicit proof that the non-local game $G_d$ defined in \eqref{eq:CGLMPd-OW} is an OW-game for all $d$.

We first write the non-local game $G_{d}$ in a way to show Alice's steering:
\begin{eqnarray}
G_{d} & = & \sum_{a=0}^{d-1}P\left(a|0\right)\sum_{\Delta=0}^{d-1}\Delta\Big[P\left(b=a-1-\Delta|a_{0},0\right)\nonumber\\
&&+P\left(b=a+1+\Delta|a_{0},1\right)\Big]\nonumber\\
 &  & +\sum_{a=0}^{d-1}P\left(a|1\right)\sum_{\Delta=0}^{d-1}\Delta\Big[P\left(b=a+1+\Delta|a_{1},0\right)\nonumber\\&&+P\left(b=a-\Delta|a_{1},1\right)\Big].
\end{eqnarray}
We assume that the maximal violation of CGLMP$_d$ can only be obtained by a suitable state and the projective measurements $ E^x_a=\left|a_{x}\right\rangle \left\langle a_{x}\right| $ and $ E^y_b=\left|b_{y}\right\rangle \left\langle b_{y}\right| $
defined by
\begin{eqnarray}
\left|a_{x}\right\rangle  & = & \frac{1}{\sqrt{d}}\sum_{k=0}^{d-1}\exp\left( i\frac{2\pi}{d}ka\right) \exp\left( ik\phi_{x}\right) \left|k\right\rangle ,\nonumber \\
\left|b_{y}\right\rangle  & = & \frac{1}{\sqrt{d}}\sum_{k=0}^{d-1}\exp\left( -i\frac{2\pi}{d}kb\right) \exp\left( ik\theta_{y}\right) \left|k\right\rangle ,\label{eq:CGLMPd-measu}
\end{eqnarray}
with $\phi_{0}=0$, $\phi_{1}=\frac{\pi}{d}$, $\theta_{0}=-\frac{\pi}{2d}$,
and $\theta_{1}=\frac{\pi}{2d}$ \cite{CGLMP}. For large $d$, this form of the optimal measurements is conjectured based on numerical results \cite{zohren2008}.

It follows that
\begin{eqnarray*}
B\left(0,a\right) & = & \sum_{\Delta=0}^{d-1}\Delta\left[E_{b=a-1-\Delta}^{y=0}+E_{b=a+1+\Delta}^{y=1}\right]\\
 & = & \sum_{k,k^{\prime}=0}^{d-1}\exp\left( -i\frac{2\pi}{d}a\left(k-k^{\prime}\right)\right) f_{0}\left(k,k^{\prime}\right)\left|k\right\rangle \left\langle k^{\prime}\right|,
\end{eqnarray*}
where $f_{0}\left(k,k^{\prime}\right)=\frac{2}{d}\sum_{\Delta=0}^{d-1}\Delta\cos\left[\frac{\pi}{2d}\left(k-k^{\prime}\right)\left(4\Delta+3\right)\right]$; and similarly,
%Similarly, for $x=1$ and different $a$, we have 
\begin{eqnarray*}
B\left(1,a\right) & = & \sum_{\Delta=0}^{d-1}\Delta\left[E_{b=a+1+\Delta}^{y=0}+E_{b=a-\Delta}^{y=1}\right]\\
 & = & \sum_{k,k^{\prime}=0}^{d-1}\exp\left( -i\frac{2\pi}{d}a\left(k-k^{\prime}\right)\right) f_{1}\left(k,k^{\prime}\right)\left|k\right\rangle \left\langle k^{\prime}\right|,
\end{eqnarray*}
where $f_{1}\left(k,k^{\prime}\right)=f_{0}\left(k,k^{\prime}\right)\exp\left(-i\frac{\pi}{d}\left(k-k^{\prime}\right)\right)$.

The four $B(x,a)$ can be transformed into each others by unitaries. For fixed $x$, $U_{a\prime a}B(x,a)U_{a^{\prime}a}^{\dagger}=B(x,a^{\prime})$ holds for $U_{a^{\prime}a}=\sum_{k}e^{-i\frac{2\pi}{d}\left(a^{\prime}-a\right)k}\left|k\right\rangle \left\langle k\right|$. Similarly, there exist $V=\sum_{k}e^{-i\frac{\pi}{d}k}\left|k\right\rangle \left\langle k\right|$ such that $B(1,a)=VB(0,a)V^{\dagger}$. This implies that all $B(x,a)$ share the same maximal eigenvalue, i.e. $\lambda_{d\left(x,a\right)}=\lambda_{d}$. We'll denote by  $\ket{\beta^{\left(x,a\right)}}$ the eigenstate associated to the maximal eigenvalue of $B\left(x,a\right)$.

Now we need to show that there always exists a bipartite pure state $\ket{\psi_{AB}}$ such that the OW criterion \eqref{owcriterion} holds. Let's set $\ket{\beta^{\left(0,0\right)}}=\sum_{k}\beta_{k}\ket{k}$; by the unitary relationship between different $B^{\left(x,a\right)}$, it follows that $\ket{\beta^{\left(0,a\right)}} =U_{a0}\ket{\beta^{\left(0,0\right)}}$ and $\ket{\beta^{\left(1,a\right)}} =U_{a0}V\ket{\beta^{\left(0,0\right)}}$. 
Using these relations, one can verify that the bipartite state $\ket{\psi_{AB}} =\sum_{k}\beta_{k}\ket{kk}$ is such that
$\ket{\psi_{B}^{\left(x,a\right)}} =\sum_{k}\exp\left(-i\frac{2\pi}{d}ak\right)\exp\left(-ik\phi_{x}\right)\beta_{k}\ket{k} \equiv \ket{\beta^{(x,a)}}$. This concludes the proof that $G_d$ is an OW-game. 
%Then we can check that $\left|\psi_{B}^{\left(0,0\right)}\right\rangle =\left|\beta^{\left(0,0\right)}\right\rangle$, $\left|\psi_{B}^{\left(0,a\right)}\right\rangle =U_{a0}\left|\beta^{\left(0,0\right)}\right\rangle =\left|\beta^{\left(0,a\right)}\right\rangle$ and $\left|\psi_{B}^{\left(1,a\right)}\right\rangle =U_{a0}U_{10}\left|\beta^{\left(0,0\right)}\right\rangle =\left|\beta^{\left(1,a\right)}\right\rangle$.

\end{appendix}

\end{document}